%% file: main.tex
\documentclass[a4paper,11pt]{article}
\pdfoutput=1 

\usepackage{jheppub} 

\usepackage[T1]{fontenc} 
\usepackage{ifpdf} 
\usepackage{flexisym}
\usepackage{breqn}
\usepackage{comment}

\usepackage{epsf,epsfig,epstopdf}
\usepackage{graphics}
\usepackage{graphicx}
\usepackage[usenames,dvipsnames]{xcolor}
\usepackage{hyperref}
\usepackage{url}
\usepackage{subfig}
\usepackage{amsmath}
\allowdisplaybreaks

\preprint{TUM-HEP-1384/21}

\title{\boldmath Two-loop non-planar hexa-box integrals with one massive leg}

\author[a]{Adam Kardos,}
\author[b]{Costas G. Papadopoulos,}
\author[c,d]{Alexander V. Smirnov,}
\author[b,e,f]{Nikolaos Syrrakos}
\author[e]{and Christopher Wever}

\affiliation[a]{University of Debrecen, Faculty of Science and Technology, Department of Experimental Physics, 4010, Debrecen, PO Box 105, Hungary}
\affiliation[b]{Institute of Nuclear and Particle Physics, NCSR Demokritos, \\Patr. Grigoriou E' \& 27 Neapoleos Str, 15341 Agia Paraskevi, Greece}
\affiliation[c]{Research Computing Center, Moscow State University, 119991 Moscow, Russia}
\affiliation[d]{Moscow Center for Fundamental and Applied Mathematics, 119992 Moscow, Russia}
\affiliation[e]{Physik-Department, Technische Universität München,\\ James-Franck-Str. 1, 85748 Garching, Germany}
\affiliation[f]{Physics Division, National Technical University of Athens, \\Zografou Campus, Athens 15780, Greece}

\emailAdd{kardos.adam@science.unideb.hu}
\emailAdd{costas.papadopoulos@cern.ch}
\emailAdd{asmirnov80@gmail.com}
\emailAdd{nikolaos.syrrakos@tum.de}
\emailAdd{christopher.wever@tum.de}

\abstract{Based on the Simplified Differential Equations approach, we present results for the two-loop non-planar hexa-box families of master integrals. We introduce a new approach to obtain the boundary terms and establish a one-dimensional integral representation of the master integrals in terms of Generalised Polylogarithms, when the alphabet contains non-factorisable square roots. The results are relevant to the study of NNLO QCD corrections for $W,Z$ and Higgs-boson production in association with two hadronic jets.}


\begin{document} 
\maketitle
\flushbottom

\input{Intro}

\input{IntegralFamilies}

\input{Boundaries}

\input{NumInt}

\input{Conclusions}

\acknowledgments
We thank the authors of reference~\cite{Abreu:2021smk} for providing their numerical results used for cross-checking our results.
\\

\noindent A.~Kardos is supported by grant K 125105 of the National Research, Development and Innovation Fund in Hungary and by
the \'UNKP-21-Bolyai+ New National Excellence Program of the Ministry
for Innovation and Technology from the source of the National Research, Development and Innovation Fund.
AK kindly acknowledges further financial support from the Bolyai Fellowship programme of the Hungarian
Academy of Sciences. The work of A.~Smirnov was supported by Russian Ministry of Science and Higher Education, agreement No. 075-15-2019-1621. N.~Syrrakos was supported by the Excellence Cluster ORIGINS funded by the Deutsche Forschungsgemeinschaft (DFG, German Research Foundation) under Germany's Excellence Strategy - EXC-2094 - 390783311. The research of C.~Wever was supported in part by the BMBF project No. 05H18WOCA1.




\bibliographystyle{JHEP}

\bibliography{hexabox.bib}

\end{document}

%% file: Intro.tex
\section{Introduction}
The computation of higher order corrections to Standard Model (SM) scattering processes and their comparison against data coming from collider experiments remains one of the best approaches for the study of Nature at its most fundamental level. The discovery of the Higgs boson at the LHC \cite{ATLAS:2012yve, CMS:2012qbp} solidified the mathematical consistency of the SM of Particle Physics as our best fundamental description of Nature. In the absence of any clear signals for physics beyond the SM, a detailed study of the properties of the Higgs boson along with a scrutinization of key SM processes have spearheaded the endeavour to advance our understanding of Particle Physics \cite{Heinrich:2020ybq}. 

The upcoming High Luminosity upgrade of the LHC will provide us with experimental data of unprecedented precision. Making sense of the data and exploiting the machine's full potential will require theoretical predictions of equally high precision. In recent years, the theoretical community has made tremendous effort to meet the challenge of performing notoriously difficult perturbative calculations in Quantum Field Theory. The current precision frontier for the QCD dominated processes studied at the LHC lies at the Next-to-Next-to-Leading-Order (NNLO) for massless $2\to3$ scattering with one off-shell external particle \cite{Amoroso:2020lgh, Gehrmann:2021qex}. 

A typical NNLO calculation involves, among other things, the computation of two-loop Feynman diagrams \cite{Tancredi:2021oiq}. The established method for performing such calculations is by solving first-order differential equations (DE) satisfied by the relevant Feynman integrals (FI) \cite{Kotikov:1990kg, Kotikov:1991hm, Kotikov:1991pm, Gehrmann:1999as}. Working within dimensional regularisation in $d=4-2\epsilon$ dimensions, allows the derivation of linear relations in the form of Integration-By-Parts (IBP) identities satisfied by these integrals \cite{Chetyrkin:1981qh}, which allows one to obtain a minimal and finite set of FI for a specific scattering process, known as master integrals (MI). 

It has been conjectured that FI with constant leading singularities in $d$ dimensions satisfy a simpler class of DE \cite{Henn:2014qga}, known as canonical DE \cite{Henn:2013pwa}. A basis of MI satisfying canonical DE is known as a pure basis. The study of the special functions which appear in the solutions of such DE has provided a deeper understanding of their mathematical properties. These special functions often admit a representation in the form of Chen iterated integrals \cite{Chen:1977oja}. For a large class of FI, their result can be written in terms of a well studied class of special functions, known as Multiple of Goncharov polylogarithms (GPLs) \cite{Goncharov:1998kja, Duhr:2011zq, Duhr:2012fh, Duhr:2014woa}.  Several computational tools have been developed for their algebraic manipulation \cite{Duhr:2019tlz} and numerical evaluation \cite{Vollinga:2004sn, Naterop:2019xaf}. 

For the case of two-loop five-point MI with one massive leg, pure bases of MI have been recently presented in \cite{Abreu:2020jxa} for the planar topologies, which we will call \textit{one-mass pentaboxes}, and more recently in \cite{Abreu:2021smk} for some of the non-planar topologies, which we will call \textit{one-mass hexaboxes}. All one-mass pentaboxes have been computed both numerically \cite{Abreu:2020jxa}, using generalised power-series expansions \cite{Moriello:2019yhu, Hidding:2020ytt}, as well as analytically in terms of GPLs \cite{Canko:2020ylt}, by employing the Simplified Differential Equations (SDE) approach \cite{Papadopoulos:2014lla}. Recently, analytic results were also obtained in the form of Chen iterated integrals and have been implemented into the so-called \textit{one-mass pentagon functions} \cite{Chicherin:2021dyp}, similar to the two-loop five-point massless results \cite{Gehrmann:2018yef, Chicherin:2020oor}. These results, along with fully analytic solutions for the relevant one-loop integral family \cite{Syrrakos:2020kba}, have lead to the production of the first phenomenological studies at the leading-colour approximation for $2\to3$ scattering processes involving one massive particle at the LHC \cite{Badger:2021nhg, Badger:2021ega, Abreu:2021asb}. For the one-mass hexabox topologies, numerical results were first presented in \cite{Papadopoulos:2019iam}, using a method which emulates the Feynman parameter technique, for one of the non-planar integral families. All three integral families were treated numerically in \cite{Abreu:2021smk} using the same methods as in \cite{Abreu:2020jxa}. 

\begin{figure}[t!]
\centering
\includegraphics[width=0.20 \linewidth]{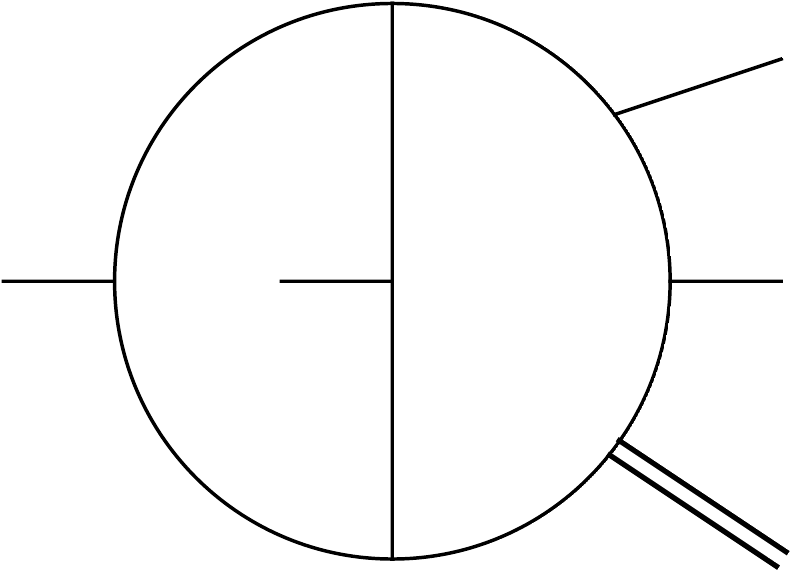} \hspace{0.6 cm}
\includegraphics[width=0.20 \linewidth]{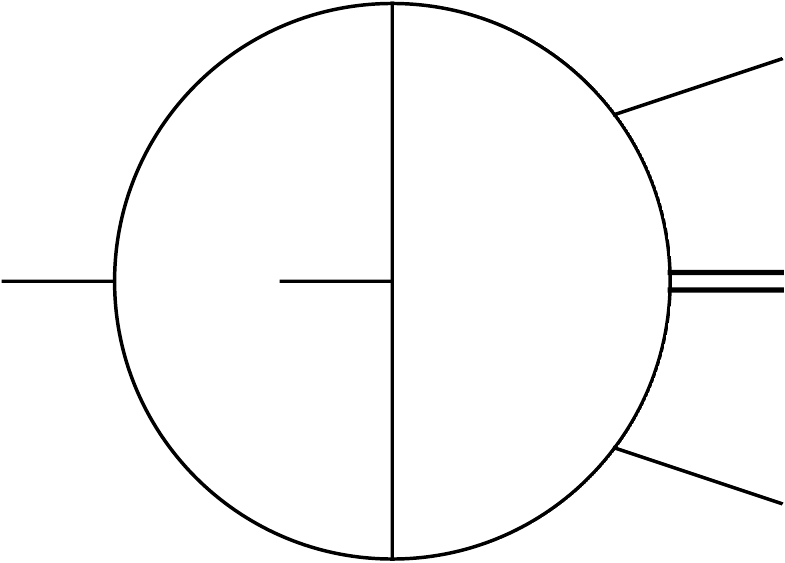} \hspace{0.6 cm}
\includegraphics[width=0.20 \linewidth]{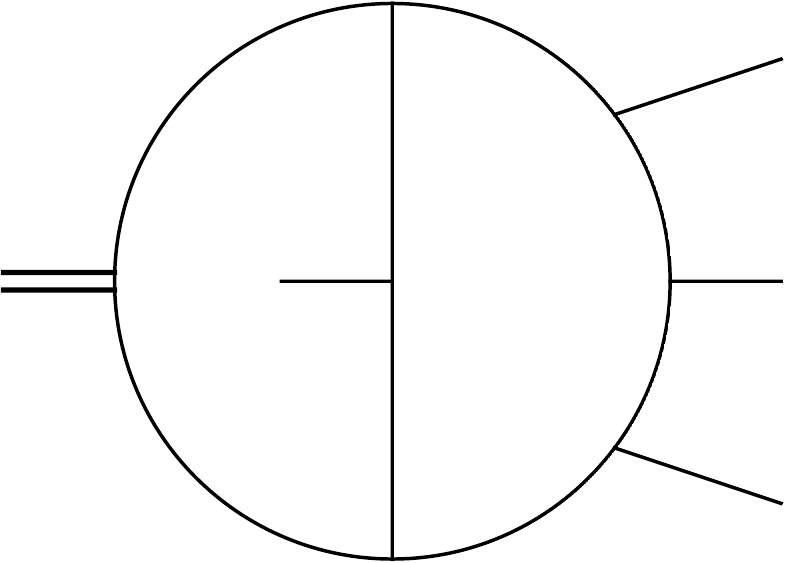} 
\\[12pt]
\includegraphics[width=0.20 \linewidth]{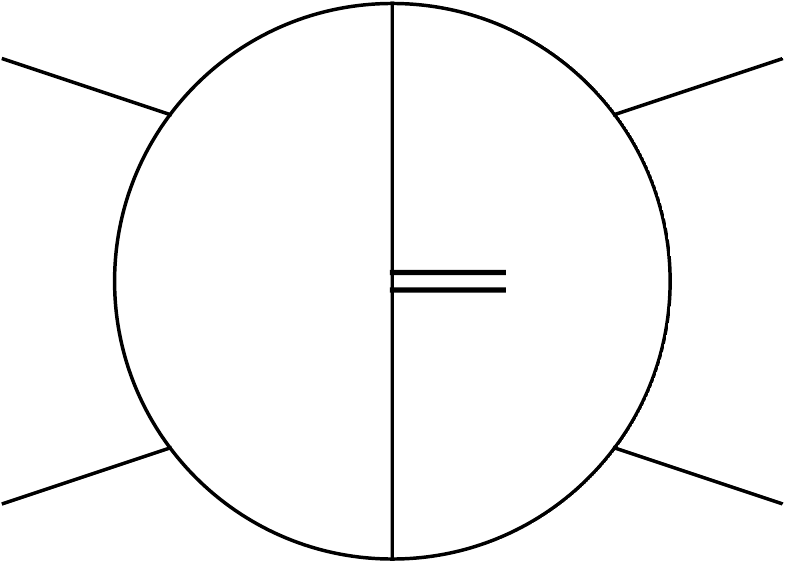} \hspace{0.6 cm}
\includegraphics[width=0.20 \linewidth]{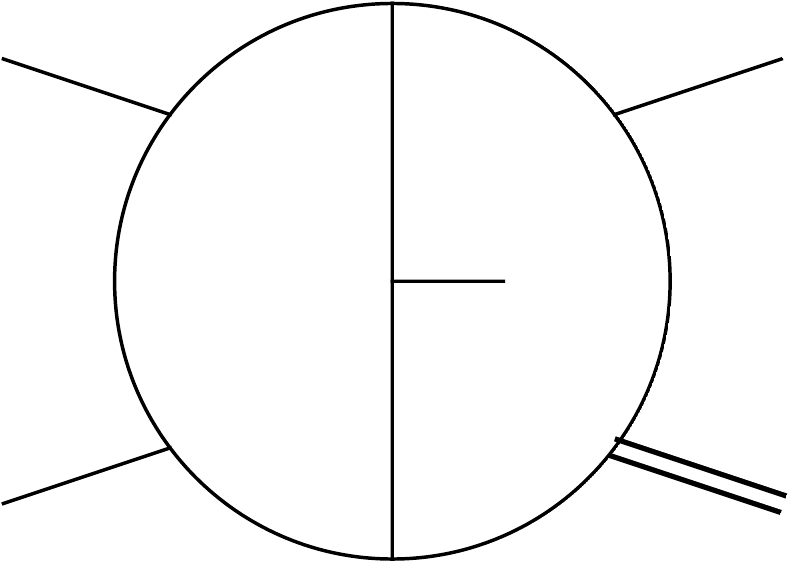}
 \caption{The five non-planar families with one external massive leg. The first row corresponds to the so-called hexabox topologies, whereas the diagrams of the second row are known as double-pentagons. We label them as follows: $N_1$ (top left), $N_2$ (top middle), $N_3$ (top right), $N_4$ (bottom left), $N_5$ (bottom right). All diagrams have been drawn using {\tt Jaxodraw} \cite{Binosi:2008ig}.}
\label{fig:fivepoint}
\end{figure}

In this paper, we employ the SDE approach and obtain semi-analytic results for all one-mass hexaboxes, using the pure bases presented in \cite{Abreu:2021smk}. More specifically, we obtain fully analytic expressions in terms of GPLs of up to weight 4 for the first non-planar family, denoted as $N_1$ in figure \ref{fig:fivepoint}. For families $N_2$ and $N_3$, we obtain analytic results for the unknown non-planar integrals up to weight 2, whereas for weights 3 and 4 we introduce a one-fold integral representation in terms of GPLs allowing for a straightforward numerical evaluation of our expressions. 

%% file: IntegralFamilies.tex
\section{Hexabox integral families}
\label{sec1}

There are three non-planar families of MI that correspond to the one-mass hexabox topologies, labelled as $N_1$, $N_2$ and $N_3$, see figure~\ref{fig:fivepoint}.
We adopt the definition of the scattering kinematics following~\cite{Abreu:2021smk}, where external momenta $q_i,\; i=1\ldots 5$ satisfy $\sum_1^5 q_i=0$, $q_1^2\equiv p_{1s}$, $q_i^2=0,\; i=2\ldots 5$, and the six independent invariants are given by $\{ q_1^2,s_{12},s_{23},s_{34},s_{45},s_{15}\}$, with $s_{ij}:=\left(q_i+q_j\right)^2$.

In the SDE approach~\cite{Papadopoulos:2014lla} the momenta are parametrized by introducing a dimensionless variable $x$, as follows 
\begin{gather}
q_1 \to p_{123}-x p_{12},\; q_2 \to p_4,\; q_3 \to -p_{1234},\; q_4 \to x p_1
\label{eq:momx}
\end{gather}
where the new momenta $p_i,\; i=1\ldots 5$ satisfy now $\sum_1^5p_i=0$, $p_i^2=0,\; i=1\ldots 5$, whereas  $p_{i\ldots j}:=p_i+\ldots +p_j$. The set of independent invariants is given  by $\{ S_{12},S_{23},S_{34},S_{45},S_{51},x\}$, with $S_{ij}:=\left(p_i+p_j\right)^2$. The explicit mapping between the two sets of invariants is given by

\begin{gather}
q_1^2=(1-x)(S_{45}-S_{12} x),
\;s_{12}=\left(S_{34} - S_{12}(1 - x)\right) x,
\;s_{23}= S_{45},
\;s_{34} = S_{51} x,
\nonumber \\
\;s_{45}=S_{12}x^2,
\;s_{15}=S_{45} + (S_{23} - S_{45}) x
    \label{eq:itatoours}
\end{gather}
and as usual the $x=1$ limit corresponds to the on-shell kinematics.

The corresponding Feynman Integrals are defined through

\begin{gather}
F^{N_1}_{a_1\cdots a_{11}}:=e^{2\gamma_E \epsilon} \int \frac{d^dk_1}{i\pi^{d/2}}\frac{d^dk_2}{i\pi^{d/2}}
\frac{1}{k_1^{2a_1} (k_1 +  q_1)^{2a_2} (k_1 +  q_{12})^{2a_3} (k_1 + q_{123})^{2a_4}} \nonumber\\
\times \frac{1}{ (k_1 + k_2 + q_{1234})^{2a_5} (k_1 + k_2)^{2a_6} k_2^{2a_7} (k_2 + q_4)^{2a_{8}}(k_2 + q_{1})^{2a_9}
(k_1 + q_4)^{2a_{10}}(k_2 + q_{12})^{2a_{11}} }
, \label{eq::N1}
\end{gather}

\begin{gather}
F^{N_2}_{a_1\cdots a_{11}}:=e^{2\gamma_E \epsilon} \int \frac{d^dk_1}{i\pi^{d/2}}\frac{d^dk_2}{i\pi^{d/2}}
\frac{1}{k_1^{2a_1} (k_1 - q_{1234})^{2a_2} (k_1 -  q_{234})^{2a_3} (k_1 - q_{34})^{2a_4}} \nonumber\\
\times \frac{1}{(k_1 + k_2 - q_4)^{2a_5} (k_1 + k_2)^{2a_6} k_2^{2a_7} 
(k_2 + q_3)^{2a_8} (k_2 - q_{1234})^{2a_{9}} (k_1 + q_{3})^{2a_{10}} (k_2 - q_{234})^{2a_{11}}}, \label{eq::N2}
\end{gather}

\begin{gather}
F^{N_3}_{a_1\cdots a_{11}}:=e^{2\gamma_E \epsilon} \int \frac{d^dk_1}{i\pi^{d/2}}\frac{d^dk_2}{i\pi^{d/2}}
\frac{1}{k_1^{2a_1} (k_1 + q_2)^{2a_2} (k_1 +  q_{23})^{2a_3} (k_1 + q_{234})^{2a_4}} \nonumber\\
\times \frac{1}{(k_1 + k_2 + q_{1234})^{2a_5} (k_1 + k_2)^{2a_6} k_2^{2a_7} 
(k_2 + q_1)^{2a_8} (k_2 + q_2)^{2a_{9}} (k_1 + q_1)^{2a_{10}} (k_2 + q_{23})^{2a_{11}}}, \label{eq::N3}
\end{gather}

where $q_{i\ldots j}:=q_i+\ldots +q_j$.

Using {\tt FIRE6}~\cite{Smirnov:2019qkx} we found that the $N_1$ family consists of 86 MI out of which 10 MI are genuinely new, the rest being known from the one-mass planar pentabox~\cite{Canko:2020ylt} or the non-planar double-box families~\cite{Papadopoulos:2014hla}. For $N_2$ and $N_3$ the corresponding numbers are 86, 13 and 135, 21.

\subsection{Pure bases and simplified canonical differential equations}
We adopt the pure bases presented in \cite{Abreu:2021smk}. As was the case for the pure bases of the planar families presented in \cite{Abreu:2020jxa}, a $\mathrm{d}\log$ form of the relevant differential equations was achieved, whose alphabet involves several square roots of the kinematic invariants $\{ q_1^2,s_{12},s_{23},s_{34},s_{45},s_{15}\}$. More specifically, the following six square roots that appear in the alphabets of the one-mass hexabox integral families are
\begin{align}
    r_1 &= \sqrt{\lambda(p_{1s},~s_{23},~s_{45})}\\
    r_2 &= \sqrt{\lambda(p_{1s},~s_{24},~s_{35})}\\
    r_3 &= \sqrt{\lambda(p_{1s},~s_{25},~s_{34})}\\
    r_4 &= \sqrt{\det\mathbb{G}(q_1,~q_2,~q_3,~q_4)}\\
    r_5 &= \sqrt{\Sigma_5^{(1)}}\\
    r_6 &= \sqrt{\Sigma_5^{(2)}}
\end{align}
with $\lambda(x,y,z) = x^2-2 x y-2 x z+y^2-2 y z+z^2$ representing the K\"allen function, $\mathbb{G}(q_1,~q_2,~q_3,~q_4) = \{2 q_i\cdot q_j\}$ being the the Gram matrix of the external momenta, and $\Sigma_5^{(1)},~\Sigma_5^{(2)}$ are the polynomials
{\small
\begin{align}
    \Sigma_5^{(1)} &= s_{12}^2 \left(s_{15}-s_{23}\right){}^2+\left(s_{23} s_{34}+\left(s_{15}-s_{34}\right) s_{45}\right){}^2 \nonumber\\
    &+2 s_{12} \left(-s_{45} s_{15}^2+s_{23} s_{34} s_{15}+\left(s_{23}+s_{34}\right) s_{45} s_{15}+s_{23} s_{34}
   \left(s_{45}-s_{23}\right)\right) \\
    \Sigma_5^{(2)} &=\left(s_{12} \left(p_{\text{1s}}-s_{15}+s_{23}\right)-s_{23} s_{34}\right){}^2+s_{45}^2 \left(p_{\text{1s}}-s_{15}+s_{34}\right){}^2\nonumber \\
    &-2 s_{45} \left(s_{34} \left(\left(s_{12}+s_{23}\right) p_{\text{1s}}-s_{15} s_{23}-s_{12}
   \left(s_{15}+s_{23}\right)\right)+s_{12} \left(p_{\text{1s}}-s_{15}\right) \left(p_{\text{1s}}-s_{15}+s_{23}\right)+s_{23} s_{34}^2\right).
\end{align}
}
For topology $N_1$, the square roots $r_1~\text{and}~r_4$ appear in its alphabet given in \cite{Abreu:2021smk}. Introducing the dimensionless variable $x$ rationalises these two roots through the mapping of \eqref{eq:itatoours}. This allows us to derive a SDE in canonical form for $N_1$,
\begin{equation}\label{eq:cansde}
\partial_{x} \textbf{g}=\epsilon \left( \sum_{i=1}^{l_{max}} \frac{\textbf{M}_i}{x-l_i} \right) \textbf{g}
\end{equation}
where $\textbf{g}$ is the pure basis of $N_1$, $\textbf{M}_i$ are the residue matrices corresponding to each letter $l_i$ and $l_{max}$ is the length of the alphabet, which for $N_1$ is $l_{max}=21$. It is interesting to note here the significant reduction in the number of letters in comparison with the alphabet of $N_1$ given in \cite{Abreu:2021smk}, where the relevant length of the alphabet is 39. The form of \eqref{eq:cansde} allows for a direct iterative solution order-by-order in $\epsilon$ in terms of GPLs, assuming that the relevant boundary terms are obtained.


For topologies $N_2$ and $N_3$, the square roots appearing in their respective alphabets \cite{Abreu:2021smk} are $\{r_1,~r_2,~r_4,~r_5\}$ and $\{r_1,~r_3,~r_4,~r_6\}$. In general all the square roots with the exception of $\{r_5,~r_6\}$ can be rationalised using either the mapping given in \eqref{eq:itatoours} or a variant of it~\cite{Canko:2020ylt,Papadopoulos:2014lla}. Nevertheless, in order to write an equation in the form of \eqref{eq:cansde} a {\it simultaneous} rationalisation of all square roots is necessary. In fact, the mapping \eqref{eq:itatoours} allows for the rationalisation of $r_1~\text{and}~r_4$ in terms of $x$, but this is not the case for $\{r_2,~r_3,~r_5,~r_6\}$. It is thus not possible to achieve a canonical SDE in the form of \eqref{eq:cansde} for families $N_2$ and $N_3$ using the parametrisation \eqref{eq:momx}. This does not mean that the basis elements cannot be cast in the form of GPLs, but just that such a representation is not straightforwardly obtained based on the simple equation \eqref{eq:cansde}. The more general form of the SDE takes the form:
\begin{equation}\label{eq:sde}
{\partial _x}{\bf{g}} = \epsilon \left( {\sum\limits_{a = 1}^{{l_{max}}} {\frac{{d{L_a}}}{{dx}}{{\bf{M}}_a}} } \right){\bf{g}}
\end{equation}
where most of the $L_a$ are simple rational functions of $x$, as in \eqref{eq:cansde}, whereas the rest are algebraic functions of $x$ involving the non-rationalisable square roots.

A detailed analysis of \eqref{eq:sde} reveals that these non-rationalisable square roots start appearing at weight two. 
In practise this means that we can use the mapping \eqref{eq:itatoours} and solve the respective canonical DE for $N_2$ and $N_3$ by integrating with respect to $x$ up to weight one in terms of ordinary logarithms. For weight two, analytic expressions in terms of GPLs can be achieved due to the fact that the non-rationalisable square roots $\{r_2,~r_3,~r_5,~r_6\}$ appear decoupled in the DE. In fact, most of the basis elements are straightforwardly expressed in terms of GPLs by integrating the corresponding DE. For the rest, an educated ansatz can be constructed involving only specific weight-two GPLs, which are identified by inspecting the DE in each case where square roots $\{r_2,~r_3,~r_5,~r_6\}$ appear, modulo the boundary terms that one needs to compute. Thus analytic expressions in terms of GPLs up to weight two are obtained for all elements belonging in these families.

To further elaborate on this point let us analyse a rather simple case of a 3-point integral sector with three off-shell legs, that appears in both $N_2$ and $N_3$ families. This sector comprises two basis elements and the DE satisfied by those elements includes also two-point MI that are known in closed form. For instance, in $N_2$, the 3-point integrals appear as basis elements number 10 and 11 (see the ancillary file). The element 10 at weight 2, $g_{10}^{\left( 2 \right)}$, can straightforwardly be obtained by integrating the \eqref{eq:sde} and it is expressible in terms of GPLs in the form ${\cal G}(a,b;x)$ where $a,b$ are independent of $x$. On the contrary the element 11 at weight 2, $g_{11}^{\left( 2 \right)}$, is obtained by construction of an ansatz. Let us mention that all elements in question, except those involving the square roots $\{r_5,~r_6\}$, namely element 73 in $N_2$ and 114 in $N_3$, are known in terms of GPLs up to weight 4~\cite{Canko:2020ylt,Papadopoulos:2014hla}, based though on different variants of the parametrization \eqref{eq:momx}. For instance element 11 of $N_2$ is given as
\begin{align}
g_{11}^{\left( 2 \right)}& = 8\Bigg(2{\cal G}(0, - y)\left( {{\cal G}\left( {1,y} \right) - {\cal G}\left( {\frac{{{{\tilde S}_{45}}}}{{{{\tilde S}_{12}}}},y} \right)} \right) + 2{\cal G}\left( {0,\frac{{{{\tilde S}_{45}}}}{{{{\tilde S}_{12}}}},y} \right) - {\cal G}\left( {1,y} \right)\log \left( {\frac{{{{\tilde S}_{45}}}}{{{{\tilde S}_{12}}}}} \right){\rm{ }}\nonumber\\
& + \log \left( {\frac{{{{\tilde S}_{45}}}}{{{{\tilde S}_{12}}}}} \right){\cal G}\left( {\frac{{{{\tilde S}_{45}}}}{{{{\tilde S}_{12}}}},y} \right) - 2{\cal G}\left( {0,1,y} \right)\Bigg)
\label{eq:g11at2}
\end{align}
where the new parametrization of the external momenta is given by 
\begin{equation}
{q_1} \to {{\tilde p}_{123}} - y{{\tilde p}_{12}},\;{q_2} \to y{{\tilde p}_2},\;{q_3} \to  - {{\tilde p}_{1234}},\;{q_4} \to y{{\tilde p}_1}
\label{eq:momy}
\end{equation}
with the new momenta ${\tilde p}_i,\; i=1\ldots 5$ satisfying as usual, $\sum_1^5{\tilde p}_i=0$, ${\tilde p}_i^2=0,\; i=1\ldots 5$, with  ${\tilde p}_{i\ldots j}:={\tilde p}_i+\ldots +{\tilde p}_j$. The set of independent invariants is given by $\{ {\tilde S}_{12},{\tilde S}_{23},{\tilde S}_{34},{\tilde S}_{45},{\tilde S}_{51},y\}$, with ${\tilde S}_{ij}:=\left({\tilde p}_i+{\tilde p}_j\right)^2$. The explicit mapping between the two sets of invariants is given by
\begin{gather}
    q_1^2 = (1 - y)({{\tilde S}_{45}} - {{\tilde S}_{12}}y),\;{s_{12}} = {{\tilde S}_{45}}(1 - y) + {{\tilde S}_{23}}y,\;{s_{23}} =  - y\left( {{{\tilde S}_{12}} - {{\tilde S}_{34}} + {{\tilde S}_{51}}} \right),\nonumber\\
    {s_{34}} = {{\tilde S}_{51}}y,{s_{45}} = y\left( {{{\tilde S}_{23}} - {{\tilde S}_{45}} - {{\tilde S}_{51}}} \right),\;{s_{15}} = y\left( {{{\tilde S}_{34}} - {{\tilde S}_{12}}(1 - y)} \right).
\label{eq:itatoy}
\end{gather}
Notice that the result of \eqref{eq:g11at2} is obtained through SDE approach in the parametrization of \eqref{eq:momy}. 
By identifying  $f_-=y$ and ${f_ + } = y\frac{{{{\tilde S}_{12}}}}{{{{\tilde S}_{45}}}}$, which in terms of \eqref{eq:itatoours} are given as 
\[{f_ \pm } = \frac{{{S_{45}} + x\left( { - {S_{23}} - {S_{34}} + 2{S_{51}} + {S_{12}}x} \right) \pm {r_2}}}{{2\left( {{S_{12}} - {S_{34}} + {S_{51}}} \right)x}}\]
we can write the DE for this element in the simple and compact form 
\[\frac{d}{{dx}}g_{11}^{\left( 2 \right)} =  - 8\left( {{\rm{dlog}}\left( {\frac{{{f_ + } - 1}}{{{f_ - } - 1}}} \right)\log \left( {{f_ - }{f_ + }} \right) - {\rm{dlog}}\left( {\frac{{{f_ + }}}{{{f_ - }}}} \right)\log \left( {\left( {{f_ - } - 1} \right)\left( {{f_ + } - 1} \right)} \right)} \right).\]
The form of the DE makes the determination of the ansatz rather straightforward, with the result 
\begin{equation}\label{eq:g11}
g_{11}^{\left( 2 \right)}=- 8\bigg( - \log ({f_ - }{f_ + })\Big({\cal G}(1,{f_ - }) - {\cal G}(1,{f_ + })\Big) + 2{\cal G}(0,1,{f_ - }) - 2{\cal G}(0,1,{f_ + })\bigg).
\end{equation}

Concerning the other non-rationalisable square root in the family $N_2$, $r_5$, it also appears for the first time at weight 2 in the basis element 73 only (see the ancillary file), which is one of the new integrals to be calculated. Following the same procedure as for the element 11, namely writing the corresponding DE in a similar form, we find that the expression at weight 2 is similar to that of \eqref{eq:g11},
\begin{equation}\label{eq:g73}
g_{73}^{\left( 2 \right)} = 16\log \left( {{f_ - }{f_ + }} \right)\Big( {{\cal G}(1,{f_ - }) - {\cal G}(1,{f_ + })} \Big) - 32\Big( {\cal G}{(0,1,{f_ - }) - {\cal G}(0,1,{f_ + })} \Big)
\end{equation}
with
\[{f_ \pm } = \frac{{{S_{45}}\left( {2{S_{12}}x - {S_{34}}x + {S_{51}}} \right) + x\left( {{S_{23}}{S_{34}} - {S_{12}}{S_{23}} + x{S_{12}}{S_{51}}} \right) \pm {r_5}}}{{2{S_{45}}\left( {{S_{12}} - {S_{34}} + {S_{51}}} \right)}}\]

Regarding family $N_3$, there are two 3-point integral sectors with three off-shell legs that involve square root $r_4$, which is not rationalised in terms of $x$ by \eqref{eq:momx}, and consist of elements 12, 13 and 16, 17. Similarly to element 11 of family $N_2$, elements 12 and 16 cannot be expressed in terms of GPLs through a straightforward integration of their respective DE. However, we can achieve a GPL representation for them at weight 2 similar to \eqref{eq:g11}, where now the $f_-, f_+$ functions involve the square root $r_4$ instead of $r_2$. Square root $r_6$ appears for the first time at weight 2 in element 114 similarly to the way square root $r_5$ appears in element 73 in the $N_2$ family, allowing us to obtain an expression at weight 2 as in \eqref{eq:g73}, with the $f_-, f_+$ functions involving $r_6$ instead of $r_5$.

Studying basis elements that are known in terms of GPLs up to weight 4, proved useful in constructing an educated ansatz for the unknown integrals at weight 2. It would be very interesting to further pursue this direction, with the aim to establish a systematic way to construct representations in terms of GPLs for weights higher than 2. This will allow to extend the SDE approach to cases where the letters $L_a$ in \eqref{eq:sde} assume a general algebraic form. 
Constructing analytic expressions in terms of GPLs beyond weight 2 
%
by applying a more general procedure following the ideas of \cite{Heller:2019gkq, Heller:2021gun} is also possible, but it requires a significant amount of resources and it might well result to a proliferation of GPLs. A more practical and direct approach, introducing a one-dimensional integral representation will be presented in detail in section \ref{sec:numint}.

%% file: Boundaries.tex
\section{Boundary terms}
\label{sec:bounds}
In this section we will describe the analytic computation of all necessary boundary terms in terms of GPLs with rational functions of the underline kinematic invariants $S_{ij}$ up to weight 4. We perform this task for all three non-planar families. 

Our main approach is the one introduced in \cite{Canko:2020ylt} and elaborated in detail in \cite{Canko:2020gqp}. In general we need to calculate the $x\to0$ limit of each pure basis element. At first we exploit the canonical SDE at the limit $x\to0$ and define through it the resummation matrix 
\begin{equation}\label{eq:resmzero}
    \textbf{R} =\textbf{S} \mathrm{e}^{\epsilon \textbf{D} \log(x)} \textbf{S}^{-1}
\end{equation}
where the matrices $\textbf{S},~ \textbf{D}$ are obtained through the Jordan decomposition of the residue matrix for the letter $l_1=0$,  $\textbf{M}_1$,
\begin{equation}\label{eq:jordandeczero}
    \textbf{M}_1 =\textbf{S} \textbf{D} \textbf{S}^{-1}.
\end{equation}
Secondly, we can relate the elements of the pure basis to a set of MI $\textbf{G}$ through IBP reduction,
\begin{equation}\label{eq:gtomasters}
    \textbf{g}=\textbf{T}\textbf{G}.
\end{equation}
Using the expansion by regions method \cite{Jantzen:2012mw} as implemented in the \texttt{asy} code which is shipped along with \texttt{FIESTA4} \cite{Smirnov:2015mct}, we can obtain the $x\to0$ limit of the MI in terms of which we express the pure basis \eqref{eq:gtomasters},
\begin{equation}\label{eq:regions}
    {G_i}\mathop  = \limits_{x \to 0} \sum\limits_j x^{b_j + a_j \epsilon }G^{(b_j + a_j \epsilon)}_{i} 
\end{equation}
where $a_j$ and $b_j$ are integers and $G_i$ are the individual members of the basis $\textbf{G}$ of MI in \eqref{eq:gtomasters}. This analysis allows us to construct the following relation
\begin{equation}\label{eq:bounds}
    \mathbf{R} \mathbf{b}=\left.\lim _{x \rightarrow 0} \mathbf{T} \mathbf{G}\right|_{\mathcal{O}\left(x^{0+a_{j} \epsilon}\right)}
\end{equation}
where the right-hand side implies that, apart from the terms $x^{a_i  \epsilon}$ coming from \eqref{eq:regions}, we expand around $x=0$, keeping only terms of order $x^0$. Equation \eqref{eq:bounds} allows us in principle to determine all boundary constants $\textbf{b}=\sum_{i=0}^{6}~ \epsilon^i~ \textbf{b}_0^{(i)}$. 

More specifically, in the case where $\textbf{D}$ in \eqref{eq:jordandeczero} is non-diagonal, we will get logarithmic terms in $x$ on the left-hand side of \eqref{eq:bounds}, in the form $x^{a_j \epsilon}\log(x)$. Since no such terms appear on the right-hand side of \eqref{eq:bounds}, a set of linear relations between elements of the array $\textbf{b}$ are obtained by setting the coefficient of $x^{a_j \epsilon}\log(x)$ terms to zero. Furthermore, powers of $x^{a_j \epsilon}$ that appear only on the left-hand side can also yield linear relations among elements of $\textbf{b}$, by setting their coefficients to zero. We shall call these two sets of relations \textit{pure}, since they are linear relations among elements of $\textbf{b}$ with rational numbers as coefficients. These pure relations account for the determination of a significant part of the two components of the boundary array. Finally for the undetermined elements of $\textbf{b}$, several region-integrals $G^{(b_j + a_j \epsilon)}_{i}$ usually need to be calculated coming from \eqref{eq:regions}.Their calculation is straightforwardly achieved either by direct integration in Feynman-parameter space and then by using \texttt{HypExp}~\cite{Huber:2005yg,Huber:2007dx} to expand the resulting $_2{F_1}$ hypergeometric functions, or in a very few cases, by Mellin-Barnes techniques using the \texttt{MB}~\cite{Czakon:2005rk, mbtools},  \texttt{MBSums}~\cite{Ochman:2015fho} and \texttt{XSummer}~\cite{Moch:2005uc} packages\footnote{The in-house \texttt{Mathematica} package \texttt{Gsuite}, that automatically process the \texttt{MBSums} output through \texttt{XSummer} is used.}. The $\textbf{b}_0^{(i)}$ terms, with $i$ indicating the corresponding weight, consist of Zeta functions $\zeta(i)$, logarithms and GPLs of weight $i$ which have as arguments rational functions of the underline kinematic variables $\{ S_{12},S_{23},S_{34},S_{45},S_{51}\}$. 

This approach was efficient enough for the determination of all boundary terms for families $N_1$ and $N_2$. Specifically for family $N_1$, where a canonical SDE can be achieved \eqref{eq:cansde}, we can write a solution in terms of GPLs up to weight 4 in the following compact form 
\begin{align}
   \label{eq:solution}
   \textbf{g}&= \epsilon^0 \textbf{b}^{(0)}_{0} + \epsilon \bigg(\sum\mathcal{G}_{a}\textbf{M}_{a}\textbf{b}^{(0)}_{0}+\textbf{b}^{(1)}_{0}\bigg) \nonumber \\
   &+ \epsilon^2 \bigg(\sum\mathcal{G}_{ab}\textbf{M}_{a}\textbf{M}_{b}\textbf{b}^{(0)}_{0}+\sum\mathcal{G}_{a}\textbf{M}_{a}\textbf{b}^{(1)}_{0}+\textbf{b}^{(2)}_{0}\bigg) \nonumber \\
   &+ \epsilon^3 \bigg(\sum\mathcal{G}_{abc}\textbf{M}_{a}\textbf{M}_{b}\textbf{M}_{c}\textbf{b}^{(0)}_{0}+\sum\mathcal{G}_{ab}\textbf{M}_{a}\textbf{M}_{b}\textbf{b}^{(1)}_{0}+\sum\mathcal{G}_{a}\textbf{M}_{a}\textbf{b}^{(2)}_{0}+\textbf{b}^{(3)}_{0}\bigg) \nonumber \\
   &+ \epsilon^4 \bigg(\sum\mathcal{G}_{abcd}\textbf{M}_{a}\textbf{M}_{b}\textbf{M}_{c}\textbf{M}_{d}\textbf{b}^{(0)}_{0}+\sum\mathcal{G}_{abc}\textbf{M}_{a}\textbf{M}_{b}\textbf{M}_{c}\textbf{b}^{(1)}_{0}\nonumber \\
   &+ \sum\mathcal{G}_{ab}\textbf{M}_{a}\textbf{M}_{b}\textbf{b}^{(2)}_{0}+\sum\mathcal{G}_{a}\textbf{M}_{a}\textbf{b}^{(3)}_{0}+\textbf{b}^{(4)}_{0}\bigg)
\end{align}
were $\mathcal{G}_{ab\ldots}:= \mathcal{G}(l_a,l_b,\ldots;x)$ represent the GPLs. These results are presented in such a way that each coefficient of $\epsilon^i$ has transcendental weight $i$. If we assign weight $-1$ to $\epsilon$, then \eqref{eq:solution} has uniform weight zero.

For family $N_3$, eq. \eqref{eq:bounds} resulted in a proliferation of region-integrals, more than 200, that one would have to calculate in order to obtain boundary terms for several higher-sector basis elements. More specifically, in order to obtain the following boundary terms 
\begin{equation}\label{eq:unknownbounds}
    \{b_{101}, b_{103}, b_{104}, b_{106}, b_{113}, b_{117}, b_{118}, b_{124}, b_{125}, b_{126}, b_{130}, b_{131}, b_{132}, b_{133}\}
\end{equation}
one would have to calculate 208 region-integrals, with 17 of them having seven Feynman parameters to be integrated, making their direct integration highly non-trivial. For all basis elements apart from \eqref{eq:unknownbounds} we were able to obtain boundary terms through \eqref{eq:bounds}.

To reduce the number of region-integrals for the computation of \eqref{eq:unknownbounds} we have investigated a different approach. The idea is rather simple and straightforward. The pure basis elements can be written in general as follows:
\begin{equation}
g = C e^{2\gamma_E \epsilon} \int \frac{d^dk_1}{i\pi^{d/2}}\frac{d^dk_2}{i\pi^{d/2}}{\frac{{P\left( {\left\{ {{D_i}} \right\},\left\{ {{S_{ij},x}} \right\}} \right)}}{{\prod\limits_{i \in \Tilde{S}}^{} {D_i^{{a_i}}} }}}
\label{element}
\end{equation}
where $D_i,\, i=1...11$, represent the inverse scalar propagators, $\Tilde{S}$ the set of indices corresponding to a given sector, $S_{ij},x$ the kinematic invariants, $P$ is a polynomial, $a_i$ are positive integers and $C$ a factor depending on $S_{ij},x$. This form is usually decomposed in terms of FI, $F_i$, 
\[g = C\sum\limits_{} {{c_i}\left( {\left\{ {{S_{ij},x}} \right\}} \right)} {F_i}\]
with $c_i$ being polynomials in $S_{ij},x$. The limit $x=0$, is then obtained, after IBP reduction, through Feynman parameter representation of the individual MI, as described in the previous paragraphs. 
An alternative approach, would be to build-up the Feynman parameter representation for the whole basis element, by considering the integral in \eqref{element} as a tensor integral and making use of the formulae from the references~\cite{Gluza:2010rn,Borowka:2014aaa}, to bring it in its Feynman parameter representation. Then, by using the expansion by regions approach~\cite{Jantzen:2012mw,Smirnov:2015mct}, we determine the regions\footnote{Only the corresponding scalar integral of \eqref{element} determines the regions.} in the limit $x=0$. Rescaling the Feynman parameters by appropriate powers of $x$, keeping the leading power in $x$, we then obtain the final result that can be written as follows:
\[b  = \sum\limits_I {{N_I}\int\limits_{}^{} {\prod\limits_{i \in {S_I}} {d{x_i}\,U_I^{{a_I}}F_I^{{b_i}}{\Pi_I}} } } \]
where $I$ runs over the set of contributing regions, $U_I$ and $F_I$ are the limits of the usual Symanzik polynomials, $\Pi_I$ is a polynomial in the Feynman parameters, $x_i$, and the kinematic invariants $S_{ij}$, and $S_I$ the subset of surviving Feynman parameters in the limit. In this way a significant reduction of the number of regions to be calculated is achieved, namely from 208 to 9. Notice that in contrast to the approach described in the previous paragraphs, only the regions $x^{-2\epsilon}$ and $x^{-4\epsilon}$ contribute to the final result, making thus the evaluation of the region-integrals simpler. Moreover, this approach overpasses the need for an IBP reduction of the basis elements in terms of MI.

%% file: NumInt.tex
\section{Integral representation}
\label{sec:numint}
After obtaining all boundary terms in section \ref{sec:bounds} and constructing analytic expressions for families $N_2$ and $N_3$ up to $\mathcal{O}(\epsilon^2)$ in terms of GPLs up to weight two, we will now introduce an one-fold integral representation for $\mathcal{O}(\epsilon^3)$ and $\mathcal{O}(\epsilon^4)$. This representation will allow us to obtain numerical results through direct numerical integration~\cite{Caron-Huot:2014lda,Gehrmann:2018yef}.

\paragraph{Weight 3:}
The differential equation \eqref{eq:sde} can be written in the form:
\begin{equation}
\partial_x g^{(3)}_{I} = \sum_a \big(\partial_x\log L_a\big)\sum_{J}c^a_{IJ}g^{(2)}_{J}    
\end{equation}
where $a$ runs over the set of contributing letters, $I,J$ run over the set of basis elements, $c^a_{IJ}$ are $\mathbb{Q}-$number coefficients read off from the matrices ${\bf{M}}_a$ and $g^{(2)}_{J}$ are the basis elements at weight 2, known in terms of GPLs. Since the lower limit of integration corresponds to $x=0$, we need to subtract the appropriate term so that the integral is explicitly finite. This is achieved as follows:
\begin{equation}\label{eq:w3}
    \partial_x g^{(3)}_{I} = \sum_a \frac{l_a}{x} \sum_{J}c^a_{IJ} g^{(2)}_{J,0} + \bigg(\sum_a \big(\partial_x\log L_a\big)\sum_{J}c^a_{IJ}g^{(2)}_{J} - \sum_a \frac{l_a}{x} \sum_{J}c^a_{IJ} g^{(2)}_{J,0}\bigg)
\end{equation}
where $g^{(2)}_{I,0}$ are obtained by expanding $g^{(2)}_{I}$ around $x=0$ and keeping terms up to order ${\cal O}\big(\log(x)^2\big)$,
and $l_a \in \mathbb{Q}$ are defined through
\begin{equation}
    \partial_x \log L_a = \frac{l_a}{x} + \mathcal{O}\big(x^0\big).
\label{eq:expansion}
\end{equation}
The DE \eqref{eq:w3} can now be integrated from $x=0$ to $x={\bar x}$, and the result is given by
\begin{equation}\label{eq:finalw3}
    g^{(3)}_{I} = g^{(3)}_{I,{\cal G}} + b^{(3)}_{I} + \int_0^{\bar{x}} \mathrm{d}x~ \bigg(\sum_a \big(\partial_x\log L_a\big)\sum_{J}c^a_{IJ}g^{(2)}_{J} - \sum_a \frac{l_a}{x} \sum_{J}c^a_{IJ} g^{(2)}_{J,0}\bigg)
\end{equation}
with $b^{(3)}_{I}$ being the boundary terms at ${\cal O}(\epsilon^3)$ and
\begin{equation}\label{eq:w3gonch}
    g_{I, {\cal G}}^{(3)}=\left.\int_{0}^{\bar{x}} \mathrm{d} x \sum_{a} \frac{l_{a}}{x} \sum_{J} c_{L J}^{a} g_{J, 0}^{(2)}\right|_{{\cal G}}
\end{equation}
with the subscript $\cal{G}$, indicating that the integral is represented in terms of GPLs (see ancillary file), following the convention
\begin{equation}\label{eq:gonch}
    \int\limits_0^{\bar x} {dx\frac{1}{x}{\cal G}\left( {\underbrace {0,...0}_n;x} \right)}  = {\cal G}\left( {\underbrace {0,...0}_{n + 1};\bar x} \right).
\end{equation}

\paragraph{Weight 4:}
At weight 4, the differential equation \eqref{eq:sde} can be written in the form:
\begin{equation}
    \partial_x g^{(4)}_{I} = \sum_a \big(\partial_x\log L_a\big)\sum_{J}c^a_{IJ}g^{(3)}_{J}
\end{equation}
which after doubly-subtracting, in order to obtain integrals that are explicitly finite as in \eqref{eq:w3}, is written as 
\begin{equation}\label{eq:w4}
     \partial_x g^{(4)}_{I}  = \sum\limits_a {{\partial _x}(\log {L_a} - L{L_a})} \sum\limits_J {c_{IJ}^a} g_J^{(3)} + \sum\limits_a {{\partial _x}} (L{L_a})\sum\limits_J {c_{IJ}^a} (g_J^{(3)} - g_{J,0}^{(3)}) + \sum\limits_a {\frac{{{l_a}}}{x}} \sum\limits_J {c_{IJ}^a} g_{J,0}^{(3)}
\end{equation}
where $LL_a$ are obtained by expanding $\log(L_a)$ around $x=0$ and keeping terms up to order ${\cal O}\big(\log(x)\big)$, and 
\begin{equation}\label{eq:g0atw=3}
    g_{I, 0}^{(3)} =g_{I, {\cal G}}^{(3)}+b_{I}^{(3)}.
\end{equation}
Now, by integrating by parts and using \eqref{eq:w3} we can write the final result as follows:
\begin{align}\label{eq:finalw4}
    g_{I}^{(4)}=&g_{I, {\cal G}}^{(4)}+b_{I}^{(4)}+\left(\sum_{a} \log L_{a} \sum_{J} c_{I J}^{a} g_{J}^{(3)}\right)-\left(\sum_{a} L L_{a} \sum_{J} c_{I J}^{a} g_{J, 0}^{(3)}\right)\nonumber \\
    &-\int_{0}^{\bar{x}} \mathrm{d} x \sum_{a}\left(\log L_{a}-L L_{a}\right) \sum_{J} c_{I J}^{a} \sum_{b} \frac{l_{b}}{x} \sum_{K} c_{J K}^{b} g_{K, 0}^{(2)} \nonumber \\
    &-\int_{0}^{\bar{x}} \mathrm{d} x \sum_{a} \log L_{a} \sum_{J} c_{I J}^{a}\left(\sum_{b}\left(\partial_{x} \log L_{b}\right) \sum_{K} c_{J K}^{b} g_{K}^{(2)}-\sum_{b} \frac{l_{b}}{x} \sum_{K} c_{J K}^{b} g_{K, 0}^{(2)}\right)
\end{align}
with $a,b$ running over the set of contributing letters, $I,J,K$ running over the set of basis elements, $b_{I}^{(4)}$ being the boundary terms at ${\cal O}(\epsilon^4)$ and
\begin{equation}\label{eq:w4gonch}
    \left.g_{I, {\cal G}}^{(4)} = \int_{0}^{\bar{x}} \mathrm{d} x\left(\sum_{a} \frac{l_{a}}{x} \sum_{J} c_{I J}^{a} g_{J, 0}^{(3)}\right)\right|_{{\cal G}}
\end{equation}
where the subscript $\cal{G}$ indicates that the integral is represented in terms of GPLs (see ancillary file), following \eqref{eq:gonch}. 

\paragraph{Implementation:}
As a proof of concept, we have implemented the final formulae \eqref{eq:finalw3} and \eqref{eq:finalw4} in {\tt Mathe\-matica}.
We use {\tt NIntegrate} to perform the one-dimensional integrals appearing in the \eqref{eq:finalw3} and \eqref{eq:finalw4}, after expressing all weight-2 functions in terms of classical polylogarithms following references~\cite{Frellesvig:2016ske}. 
For kinematic configurations where there are no singularities in the domain of integration $\left( {0,\bar x} \right)$, we have checked the new basis elements obtained by us against numerical results provided by the authors\footnote{We thank the authors of reference~\cite{Abreu:2021smk} for communicating results for a Euclidean point provided by us, not included in their publication, for $N_2$ and $N_3$ families.} of reference~\cite{Abreu:2021smk} and found full agreement.
For kinematic configurations with singularities in the domain of integration, we use the  $i\epsilon-$prescription as explained in references~\cite{Papadopoulos:2014hla,Canko:2020ylt}, as well as the convention concerning the square roots appearing in the alphabet and the basis elements as detailed in section 6.2 of reference~\cite{Abreu:2020jxa}. We provide proof-of-concept codes in the ancillary files for both the Euclidean point mentioned above as well as the first physical phase-space point of Eq.~(6.15) in reference~\cite{Abreu:2021smk}.
The reader can easily assess the performance of this straightforward implementation by running the provided codes and look at the minimum number of digits in agreement with the high-precision results from reference~\cite{Abreu:2021smk}, as well as at the number of integrand evaluations performed by  {\tt NIntegrate}. Notice that the integrand expressions involve logarithms and classical polylogarithms $\mathrm{Li}_2$ that are evaluated using very little CPU time. The parts of the formulae \eqref{eq:finalw3} and \eqref{eq:finalw4} that can be represented in terms of GPLs up to weight four, as well as the results for the $N_1$ family, for which we have all basis elements in terms of GPLs up to weight four, are evaluated with {\tt GiNaC}~\cite{Vollinga:2004sn,Vollinga:2005pk} as implemented in {\tt PolyLogTools}~\cite{Duhr:2019tlz}. In the current implementation we use the default parameters for {\tt GiNaC} and the default parameters for {\tt NIntegrate} with the exception of {\tt WorkingPrecision} and {\tt PrecisionGoal}, in order to obtain reasonable results within reasonable time, taking into account that the provided implementation serves merely as a demonstration of the correctness of our representations. For the Euclidean point the precision is typically of the order of 32 digits, which is compatible with {\tt GiNaC} setup. For the physical point, the typical precision is of the order of 25 digits, which is compatible with the expected one taking into account the numerical value of the infinitesimal imaginary part assigned to the kinematical invariants. We plan to address all the details regarding the numerical evaluation in a forthcoming publication, where an optimised implementation for all two-loop five-point basis elements based on our analytic results, in line with references~\cite{Chicherin:2020oor,Chicherin:2021dyp}, will be presented.


%% file: Conclusions.tex
\section{Conclusions}
The frontier of precision calculations at NNLO currently concerns $2\to3$ scattering process involving massless propagators and one massive external particle. At the level of FI, all planar two-loop MI have been recently computed through the solution of canonical DE both numerically~\cite{Abreu:2020jxa}, via generalised power series expansions, and analytically in terms of GPLs up to weight 4~\cite{Canko:2020ylt}, using the SDE approach~\cite{Papadopoulos:2014lla}. More recently, results in terms of Chen iterated integrals were presented and implemented in the so-called pentagon functions~\cite{Chicherin:2021dyp}.

Concerning the two-loop non-planar topologies, these can be classified into the three so-called hexabox topologies and two so-called double-pentagons, see figure \ref{fig:fivepoint}. One of the hexabox topologies, denoted as $N_1$ in figure \ref{fig:fivepoint}, was calculated numerically a few years ago using an approach which introduces a Feynman parameter and uses analytic results for the sub-topologies that are involved~\cite{Papadopoulos:2019iam}. More recently, pure bases for the three hexabox topologies satisfying DE in $\mathrm{d}\log$ form were presented in reference~\cite{Abreu:2021smk} and solved numerically using the same methods as in~\cite{Abreu:2020jxa}.

In this paper we addressed the calculation of the three two-loop hexabox topologies, $N_1,~N_2,~N_3$ in figure \ref{fig:fivepoint}, using the SDE approach. For the $N_1$ family results up to weight 4 in terms of GPLs are obtained. For the $N_2$ and $N_3$ families we have established an one-dimensional integral representation involving up to weight-2 GPLs. This allows to extend the scope of the SDE approach when non-factorisable square roots appear in the alphabet. We have also introduced a new approach to compute the boundary terms directly for the basis elements, that significantly reduces the complexity of the problem. With these new developments, we hope to complete the full set of five-point one-mass two-loop MI families in the near future and provide a solid implementation for their numerical evaluation.